\documentstyle[pra,aps]{revtex}

\newtheorem{theorem}{Theorem}

\newtheorem{lemma}{Lemma}
\newtheorem{proposition}{Proposition}

\begin{document}
\draft
\title{The geometry of entanglement witnesses and  local detection of entanglement}
\author{Arthur O. Pittenger and Morton H. Rubin} 
\address{Department of Mathematics and Statistics and} 
\address{Department of Physics} 
\address{University of Maryland, Baltimore County,
Baltimore, MD 21250}
\date{July 3, 2002}
\maketitle

\begin{abstract}
Let $H^{\left[ N\right] }=H^{\left[ d_{1}\right] }\otimes \cdots \otimes H^{\left[
d_{n}\right] }$ be a tensor product of Hilbert spaces and  
let $\tau_{0}$ be the closest separable state in the Hilbert-Schmidt 
norm to an entangled state $\rho_{0}$.  Let $\tilde{\tau}_{0}$ denote 
the closest separable state to $\rho_{0}$ along the line segment 
from $I/N$ to $\rho_{0}$ where $I$ is the identity matrix. Following 
\cite{pitrubmat} a witness $W_{0}$ detecting the entanglement of 
$\rho_{0}$ can be constructed in terms of $I, \tau_{0}$ and 
$\tilde{\tau}_{0}$. If representations of $\tau_{0}$ and 
$\tilde{\tau}_{0}$ as convex combinations of separable projections are 
known, then the entanglement of $\rho_{0}$ can be detected by local 
measurements.  G\"{u}hne \textit{et. al.} in \cite{bruss1} obtain the minimum number of measurement
settings required for a class of two qubit states. We use our geometric 
approach to generalize their result to the corresponding two qudit case 
when $d$ is prime and obtain the minimum number of measurement 
settings. In those particular bipartite cases, 
$\tau_{0}=\tilde{\tau}_{0}$. We 
illustrate our general approach with a two parameter family of three qubit 
bound entangled states for which $\tau_{0} \neq \tilde{\tau}_{0}$ and 
we show our approach works for $n$ qubits.

In \cite{pitt} we elaborated on the role of a ``far face'' of the 
separable states relative to a bound entangled state $\rho_{0}$ constructed from 
an orthogonal unextendible product base. In this paper the geometric 
approach leads to an entanglement witness expressible in terms of a 
constant times $I$ and a separable density $\mu_{0}$ on the far face 
from $\rho_{0}$. Up to a normalization this coincides with the 
witness obtained in \cite{bruss1} for the particular example analyzed
there.
\end{abstract}

\section{Motivation and notation}

An important question for quantum information theory is how to 
determine if a given state is entangled. Physically, one would like to 
do this using local measurements and classical communications. Testing 
for entanglement is closely related
to Bell inequalities \cite{bell} and subsequent elaborations of Bell's
inequalities \cite{terhal0}. Recently other tests have been suggested, such as that in \cite{ekhor}
 which relies on the theory of positive operators
and on eigenvalue estimation.

An alternate approach, which is experimentally realizable, is to define local
correlated measurements motivated by some knowledge of the structure of $\rho $ 
itself, and this approach has been elaborated in \cite{bruss1}. To
describe the problem, we first define the mathematical context. Specifically,
we assume we are working with n distinct systems so that  $\rho $ is
represented as an $N\times N$ density operating on the tensor product
Hilbert space $H^{\left[ N\right] }=H^{\left[ d_{1}\right] }\otimes \cdots
\otimes H^{\left[ d_{n}\right] }$. The set $D$ of such $N\times N$ densities
operating on $H^{\left[ N\right] }$ is a compact convex subset of the real
Hilbert space $M$ of $N\times N$ Hermitian matrices where the inner product
is defined by $\left\langle A,B\right\rangle =Tr\left[ A^{\dagger }B\right] $%
. (Since the matrices are assumed to be Hermitian, the notation ``$^{\dagger
}$'' denoting the Hermitian conjugate appears to be redundant. However, we
will have occasion to use the inner product for more general matrices.) The
set of separable densities $S$ is defined as the convex hull of the
separable projections $\pi _{1}\otimes \cdots \otimes \pi _{n}$, where
 $\pi_{k}$ is a projection on $H^{[d_{k}]}$. Since $S$ is
a compact convex subset of $D$ one can test for entanglement by showing $\rho$ 
is separated from $S$ by
a hyperplane in $M$ \cite{horodeckis3}. Geometrically the idea is clear.
Mathematically it reduces to finding a Hermitian matrix $W$ with the
property that $Tr\left( W\rho \right) $ $<0\leq $ $Tr\left( W\sigma \right) $
for every density $\sigma $ in $S$. The existence of such a $W$ is
guaranteed by the general theory of convex sets in Hilbert spaces, and $W$
is known in the quantum information literature as an ``entanglement
witness''. A nice introduction to the subject and an overview of some of the
literature can be found in \cite{terhal}.

In the context of two qubits G\"{u}hne \textit{et. al.} in \cite{bruss1} assume the
general form of a two parameter family of densities $\rho $ which 
includes a
maximally entangled state $\rho _{0}$. They construct an entanglement
witness using the eigenvector of the partial transpose of $\rho $ with the
minimal (negative) eigenvalue and find that the resulting witness does not
depend on either of the parameters. Since the separating hyperplane contains
a face of the separable states, it is \textit{optimal} in the sense that no
witness detects a strictly larger set of entangled states. (See \cite{lewenstein} 
for the definitions and \cite{pitrubmat} for an exposition
related to the approach used in this paper.)

In \cite{pitrubmat}, the authors showed how an entanglement witness $W_{0}$
sensing an inseparable $\rho _{0}$ can be constructed if one also knows the
nearest separable state $\tau _{0}$: 
\[
\left\| \rho _{0}-\tau _{0}\right\| = \inf \left\{ \left\| \rho
_{0}-\sigma \right\| :\sigma \in S\right\} .
\]
Since the norm is a continuous function and the set of separable densities
is compact, $\tau _{0}$ exists, although actually computing it is not an
easy problem in general. The entanglement witness is defined by  
\begin{equation}
W_{0}=\tau _{0}+c_{0}I-\rho _{0},
\label{entwitdef}
\end{equation}
where $I$ is the $N\times N$ identity matrix and 
\[
c_{0}=Tr\left( \tau _{0}\left( \rho _{0}-\tau _{0}\right) \right) .
\]
Details and examples of this construction are given in \cite{pitrubmat}
where it is shown that $W_{0}$ is linked to the geometry via the induced inner
product 
\begin{equation}
\left\langle \left( \rho _{0}-\tau _{0}\right) ,\left( \rho -\tau
_{0}\right) \right\rangle \equiv Tr\left( \left( \rho _{0}-\tau _{0}\right)
\left( \rho -\tau _{0}\right) \right) =-Tr\left( W_{0}\rho \right) .
\label{innerproduct}
\end{equation}
In particular the separating hyperplane contains the ``nearest'' face of $S$
consisting of separable states $\sigma $ such that $\sigma -\tau _{0}$ is
orthogonal to $\rho _{0}-\tau _{0}$. Equation (\ref{innerproduct}) can be
used to show that the extreme separable projections in the convex
representation of $\tau _{0}$ must lie in the hyperplane, and that if any
separable $\sigma $ in the nearest face has full rank then $W_{0}$ is optimal.

It was first shown in \cite{zuk} that there is a neighborhood of the
normalized identity or completely random state, $D_{0}=(1/N)I$, in which 
every state is separable . 
Given that fact, it
follows from another compactness argument that there is a nearest separable
density to $\rho _{0}$ along the line segment $[D_{0}, \rho_{0}]$:  
\begin{equation}
\tilde{\tau}_{0}=\left( 1-s_{0}\right) D_{0}+s_{0}\rho _{0}
\label{tautilde}
\end{equation}
with $0<s_{0}<1$. While $\tau_{0}$ and $\tilde{\tau}_{0}$ differ in 
general, in certain examples they are the same which simplifies the 
analysis. Thus we have the following general result.

\begin{theorem}
Suppose $\rho _{0}$ is inseparable. Using the notation above, the Hermitian
matrix 
\[
W_{0}=I\left( c_{0}+\frac{1-s_{0}}{Ns_{0}}\right) +\tau _{0}-\frac{1}{s_{0}}%
\tilde{\tau}_{0}
\]
is an entanglement witness for $\rho _{0}$ and is optimal if the nearest
face contains a separable density of full rank.  $\Box $
\end{theorem}

Thus if one knew the convex representations of $\tau _{0}$ and $\tilde{\tau}%
_{0}$ in terms of tensor products of local projections, one could define
specific coordinated local measurements that would experimentally detect the
entanglement of $\rho _{0}$ via $Tr\left( \rho _{0}W_{0}\right) $. 
Finding $\tau_{0}$ and $\tilde{\tau}_{0}$ is in general difficult but 
 can be done in a variety of special cases. The 
examples we present include those
analyzed in \cite{bruss1} as well as a two parameter family of three
qubit bound entangled densities for which $\tau _{0}$ and $\tilde{\tau}_{0}$
differ. (A bound entangled state is entangled but has positive partial 
transposes.) 

Another result in \cite{bruss1} is that three sets of coordinated local
measurements is the minimum number required in their two qubit context and
an explicit representation of the three measurements was given. It was also
asserted that at least $d+1$ such measurements would be required for a
corresponding $d\times d$ system, but no suggestion for achieving 
that bound
was provided. We show how the geometric approach to entanglement witnesses
provides a unifying theme and leads to a concrete construction for the 
$d\times d$ case when $d$ is prime. 

\section{Two qubits}

In the two qubit case, the use of the nearest separable density clarifies
some of the methodology and suggests the generalization to the $d\times d$
case. Following ref. \cite{bruss1} we take 
\begin{equation}
\rho =p\rho _{a}+\left( 1-p\right) \sigma .  \label{rhogen}
\end{equation}
$\rho _{a}$ is defined by the state $a\left| 00\right\rangle +b\left|
11\right\rangle $, where $a$ and $b$ are real with $a^{2}+b^{2}=1$, $p$ is a
parameter between $0$ and $1$, and $\sigma $ is a density close to the
normalized identity, $\left\| \sigma -D_{0}\right\|
<\delta $. The density $\sigma$ represents noise that is close to 
$D_{0}$, the completely random state. The idea is to define a separating hyperplane $W_{0}$ 
based on $\rho _{0}=\frac{1}{2}\left( \left| 00\right\rangle +\left| 11\right\rangle
\right) \left( \left\langle 00\right| +\left\langle 11\right| \right) $ and
investigate what inseparable states $\rho $ are detected by $W_{0}$. 

It has been shown in a number of places that the closest separable state to 
$\rho _{0}$ is 
\begin{equation}
\tau _{0}=\frac{2}{3}D_{0}+\frac{1}{3}\rho _{0},  
\label{tau0def}
\end{equation}
so that the roles of $\tau _{0}$ and $\tilde{\tau}_{0}$ coincide.
(References and details are given in \cite{pitrubmat}.) One computes 
$c_{0}=Tr\left( \tau _{0}\left( \rho _{0}-\tau _{0}\right) \right) =\frac{1}{6%
}$, and then eq. (\ref{entwitdef}) gives 
\begin{equation}
W_{0}=\frac{1}{3}\left( 
\begin{array}{cccc}
0 & 0 & 0 & -1 \\ 
0 & 1 & 0 & 0 \\ 
0 & 0 & 1 & 0 \\ 
-1 & 0 & 0 & 0
\end{array}
\right) .
\label{W0}
\end{equation}
This differs from the optimal witness found in \cite{bruss1} only because of the use of a
different Bell state and is a special case of the general theorem above. 

As an application we have the following result.

\begin{lemma}
A sufficient condition that $\rho=p\rho_{a}+(1-p)\sigma$, where 
$||\sigma-D_{0}|| < \delta $, is not separable is that 
\[
\left( \frac{1+4\delta }{4ab+1+4\delta }\right) <p.
\]
\end{lemma}

\textit{Proof}: 
\begin{eqnarray*}
Tr\left( W_{0}\rho \right) &=&pTr\left( W_{0}\rho _{a}\right) +\left( \left(
1-p\right) \right) Tr\left( W_{0}D_{0}\right) +\left( 1-p\right) Tr\left(
W_{0}\left( \sigma -D_{0}\right) \right) \\
&=&\frac{-2abp}{3}+\frac{1}{6}\left( 1-p\right) +\left( 1-p\right) Tr\left(
W_{0}\left( \sigma -D_{0}\right) \right) \\
&\leq &\frac{-2abp}{3}+\frac{1}{6}\left( 1-p\right) +\left( 1-p\right) \frac{%
2\delta }{3}
\end{eqnarray*}
where we have used the Cauchy-Schwarz inequality in the last step. Setting
the final expression to be less than $0$, we obtain the desired inequality.
Note that if $\delta =0$ and $a=1/\sqrt{2}$ we obtain the well known
sufficient condition $1/3<p$ for inseparability of $p\rho _{0}+\left(
1-p\right) D_{0}$. $\Box $

Having defined $W_{0}$ we need to show that the measurement can be effected
by three types of coordinated local measurements. We combine 
eq.(\ref{entwitdef})
 and eq.(\ref{tau0def}) to obtain 
\begin{equation}
W_{0}=\frac{2}{3}I-2\tau _{0}  \label{qubitw}
\end{equation}
and then use the representation of $\tau _{0}$ as a convex combination of
six separable extreme points in the face of the states of $S$ in the
separating hyperplane: 
\begin{eqnarray} 
\tau_{0}&=&\frac{1}{6}[ \left( \frac{\sigma _{0}+\sigma _{z}}{2}\otimes \frac{\sigma 
_{0}+\sigma _{z}}{2}
\right) +\left( \frac{\sigma _{0}-\sigma _{z}}{2}\otimes
\frac{\sigma_{0}-\sigma _{z}}{2}\right) 
  \nonumber \\
&& +\left(
\frac{\sigma _{0}+\sigma _{x}}{2}
\otimes\frac{\sigma_{0}+\sigma_{x}}{2}\right)
 +\left( \frac{\sigma _{0}-\sigma _{x}}{2}\otimes 
 \frac{\sigma _{0}-\sigma _{x}}{2}\right) \label{tau0rep} \\  
&& +\left( \frac{\sigma _{0}+\sigma _{y}}{2}\otimes \frac{\sigma _{0}-\sigma%
_{y}}{2}\right) +\left( \frac{\sigma _{0}-\sigma _{y}}{2}\otimes%
\frac{\sigma _{0}+\sigma _{y}}{2}\right)]
 \nonumber
\end{eqnarray}
Thus one takes coordinated local measurements along the $x,$ $y$, and $z$
axes of the Bloch sphere to compute $Tr\left( W_{0}\rho \right) =\frac{2}{3}%
-2Tr\left( \tau _{0}\rho \right) $. As shown in \cite{bruss1}, this is the
minimal number of coordinated local measurements which are required.

\section{The $\MakeLowercase{d}\times \MakeLowercase{d}$ case.}

The approach used above immediately generalizes to the bipartite $d\times d$
case when $d$ is prime ($d \neq 2$): we take an entangled ``base'' state $\rho _{0}$ for
which we can compute the nearest separable state $\tau _{0}$ and thus $W_{0}$.
We again consider the family of densities $\rho =p\rho _{a}+\left(
1-p\right) \sigma $, where $\sigma $ is close to the state $D_{0}$,
 and define
\begin{displaymath} 
\rho _{a}=\left| \psi _{a}\right\rangle
\left\langle \psi _{a}\right| \quad {\textstyle where}\quad
\left| \psi _{a}\right\rangle
=\sum\limits_{k=0}^{d-1}a_{k}\left| kk\right\rangle 
\end{displaymath}
 with real $a_{k}$ such that $\sum_{k}a_{k}^{2}=1$. 
 $\rho _{0}$ is the state with $a_{k}=1/\sqrt{d}$, and 
 \begin{equation} 
 \tau _{0}=\frac{d}{d+1}D_{0}+\frac{1}{d+1}\rho _{0}
 \label{dtau}
 \end{equation}  
is the closest separable state. (See \cite
{pitrub3} for the general result and references.) Again, $\tau_{0}$ coincides with $\tilde{\tau}_{0}$,
 simplifying the problem \cite{pitrubmat,steiner1,steiner2,witru}. From eq.(\ref{entwitdef}) 
the optimal witness for $\rho _{0}$ is $W_{0}=\frac{2}{1+d}I-d\tau _{0}$,
where $c_{0}=\frac{d-1}{d\left( d+1\right) }.$ The problem now reduces to finding
analogues of the Pauli matrices which can be used to represent $\tau _{0}$
as an appropriate convex combination of projections, as in eq.( \ref{tau0rep}).
Fortunately that analysis already has been done.

In ref. \cite{pitrub1} the authors observed that the (real) Pauli matrices can be
viewed as discrete Fourier transforms of four ``computational'' basis
matrices. Using an analogous basis for $d\times d$ matrices and the
corresponding discrete Fourier transform, one is able to define $d^{2}$
orthogonal unitary matrices 
\[
U_{d}\equiv \left\{ S_{u}:u=(j,k), \quad  0\leq j,k<d\right\} 
\]
where $S_{e}=S_{\left( 0,0\right) }$ is the $d\times d$ identity. (These
same matrices had been derived independently and in a different manner by
Fivel \cite{fivel} who used them in a study of Hamiltonians on a discrete
state space. He also derived several of the properties we include below.) As
with the two qubit case, one can define sets of tensor products of
projections, and it turns out that $\tau _{0}$ can be written as a convex
combination of $d+1$ such sets in strict analogy with the representation in 
eq.(\ref{tau0rep}). These $d+1$ sets of projections correspond to the
coordinated local measurements required in \cite{bruss1} for local detection
of entanglement of $d\times d$ states.

We briefly summarize the necessary properties of these $d$-level ``spin''
matrices and relegate proofs to the Appendix. By definition 
\[
S_{\left( j,k\right) }=\sum_{r=0}^{d-1}\eta ^{jr}\left| r\right\rangle
\left\langle r+k\right| 
\]
where addition is modulo $d$ and $\eta =\exp \left( 2\pi i/d\right) $. 
\[
Tr\left[ S_{\left( j_{1},k_{1}\right) }^{\dagger }S_{\left(
j_{2},k_{2}\right) }\right] =d\delta_{j_{1},j_{2}}
\delta_{ k_{1},k_{2}} 
\]
expresses orthogonality, and thus $U_{d}$ is a basis for $d\times d$
matrices.

Using tensor products of the $U_{d}$ spin matrices, we find that for prime $d$ 
\begin{equation}
\tau _{0}=\frac{1}{d+1}\left[ \frac{1}{d^{2}}\sum_{k=0}^{d-1}S_{k(d-k),00}+%
\sum_{j=0}^{d-1}\left( \frac{1}{d^{2}}\sum_{k=0}^{d-1}S_{(kj)(kd-kj),kk}%
\right) \right], \label{tau0drep}
\end{equation}
where $S_{ij,kl}=S_{i,k}\otimes S_{j,l}$, and it remains to show that each
of the $k$-summations can be written as a sum of tensor products of complete
sets of projections. When $d$ is odd and $u=\left( j,k\right) \neq \left(
0,0\right) $, 
\[
P_{u}\left( r\right) \equiv \frac{1}{d}\sum_{m=0}^{d-1}\left( \eta
^{r}S_{u}\right) ^{m} 
\]
is a (Hermitian) projection, and $\left\{ P_{u}\left( r\right) :\text{ }%
0\leq r<d\right\} $ is a complete set of orthogonal projections. If $%
u_{j}=(j,1)$ and $v_{j}=\left( d-j,1\right) $ for $0\leq j<d$, then
(surpressing the subscript on $u_{j}$ and $v_{j}$) 
\[
\frac{1}{d^{2}}\sum_{k=0}^{d-1}S_{(kj)(kd-kj),kk}=\frac{1}{d}
\sum_{r=0}^{d-1}P_{u}\left( r\right) \otimes P_{v}\left( d-r\right) . 
\]
The first summation (\ref{tau0drep}) has an analogous representation if $%
u=\left( 1,0\right) $ and $v=\left( d-1,0\right) $.

This completes the proof: the entanglement witness $W_{0}$ can be realized
in terms of the identity and a separable density which in turn can be
written as convex combination of $d+1$ sums of tensor products of complete
(local) projections. This attains the lower bound for the number of 
coordinated local measurements as asserted in \cite{bruss1}.

As in the two qubit case, the entanglement witness $W_{0}$ detects
entanglement for a range of densities of the form $\rho =p\rho _{a}+\left(
1-p\right) \sigma $. The computation is similar to that for Lemma 1, and we
omit the details.

\begin{lemma}
$\rho =p\rho _{a}+\left( 1-p\right) \sigma $ is inseparable provided 
\[
\frac{1-p}{p}\left( 1-\frac{1}{d}+\delta \sqrt{2d\left( d-1\right) }\right)
<\left( \sum_{k=0}^{d-1}a_{k}\right) ^{2}-1
\]
where $\left\| \sigma -D_{0}\right\| <\delta $. When $d=2$, this reduces to
the inequality in Lemma 1.\ $\Box $
\end{lemma}

\section{A three qubit example}

The geometric appproach also works for a particular two parameter family of
three qubits which have positive partial transforms but are inseparable.
Since these densities are not generated by complete \textit{UPB} sets, it is
not clear that other techniques can be used to define an appropriate
entanglement witness.

Let $-1/8\leq c,d\leq 1/8$ and define the three qubit density matrix
\[
\rho(c,d)=\left( 
\begin{array}{cccccccc}
1/8 & 0 & 0 & 0 & 0 & 0 & 0 & 1/8 \\ 
0 & 1/8 & 0 & 0 & 0 & 0 & 1/8 & 0 \\ 
0 & 0 & 1/8 & 0 & 0 & c & 0 & 0 \\ 
0 & 0 & 0 & 1/8 & d & 0 & 0 & 0 \\ 
0 & 0 & 0 & d & 1/8 & 0 & 0 & 0 \\ 
0 & 0 & c & 0 & 0 & 1/8 & 0 & 0 \\ 
0 & 1/8 & 0 & 0 & 0 & 0 & 1/8 & 0 \\ 
1/8 & 0 & 0 & 0 & 0 & 0 & 0 & 1/8
\end{array}
\right) 
\]
It is convenient to identify $\rho \left( c,d\right) $ with the four vector 
$\left\langle d,c,1/8,1/8\right\rangle $ defined by the negative 
diagonal. We will use this notation for
densities with analogous structure. Further, it simplifies calculations to
use $m=\left( c+d\right) /2$ and $t=\left( c-d\right) /2$,\ and we
abuse notation by writing $\rho \left( m,t\right) $ for the same
density and $\left\langle m-t,m+t,1/8,1/8\right\rangle $ for its four
vector. The following result is proved in \cite{pitrub2} for analogous 
densities $\rho(c,d)$ for $n$ qubits defined by the $2^{n-1}$ vector 
with equal numbers of c's and d's and $2^{n-2}$ entries of 
$1/2^{n}$, $\langle 
d,\cdots,d,c,\cdots,c,1/2^{n},\cdots,1/2^{n}\rangle$.
\begin{proposition}
$\rho \left( c,d\right) $ has positive partial transposes and is completely
separable if and only if $c=d$ ($t=0$).\ $\Box $
\end{proposition}

In the $d\times d$ cases analyzed above, the line segment from $D_{0}$ to $%
\rho _{0}$ was orthogonal to the nearest separable face, and that property
characterized $\tau _{0}$, the nearest separable density to $\rho _{0}$.
Unfortunately, as shown in \cite{pitrubmat}, that perpendicularity is lost when one goes to three systems 
 and the nearest separable state $\tilde{\tau}_{0}$ to 
$\rho _{0}$ on the line segment $[D_{0},\rho _{0}]$ does not coincide with 
the closest separable state $\tau _{0}$. 
However, one can still take advantage of the geometry \textit{provided} 
$\tilde{\tau}_{0}$ lies in the nearest separable face to $\rho _{0}$.

To pursue this idea for $\rho \left( c,d\right) $, we need some additional
notation. Without loss of generality we take $c>d$ so that $t>0$, and
 let $\sigma _{1}=\sigma _{x}$ and $\sigma _{2}=\sigma _{y}$. Let $\sigma
_{0}$ denote the $2\times 2$ identity and define 
\begin{equation}
P_{jkl}^{\pm }=\frac{1}{8}\left[ \sigma _{0}\otimes \sigma _{0}\otimes
\sigma _{0}\pm \sigma _{j}\otimes \sigma _{k}\otimes \sigma _{l}\right]
\label{Pjkl}
\end{equation}
where \textit{j}, \textit{k}, and \textit{l} will take the values $1$ or $2$.
It is an easy exercise to represent such a $P_{jkl}^{\pm }$ as an average
of four projections and to confirm that  
\begin{eqnarray*}
\rho \left( m,t\right) &=&\left( \frac{1}{2}+4m\right) P_{111}^{+}+\left( 
\frac{1}{2}-4m\right) P_{221}^{-} \\
&&+4t\left( P_{212}^{-}+P_{122}^{+}\right) -8tD_{0}.
\end{eqnarray*}

As it happens, a study of the $m=0$ case is key to the analysis, and we take
as a candidate for $\tilde{\tau}_{0}\left( 0,t\right) $ the normalization of
the first part of $\rho \left( 0,t\right) $: 
\begin{equation}
\tilde{\tau}_{0}\left( 0,t\right) =\frac{1}{1+8t}\left[ \frac{1}{2}\left(
P_{111}^{+}+P_{221}^{-}\right) + 4t\left( P_{212}^{-}+P_{122}^{+}\right)
\right] .  \label{tau0tilda}
\end{equation}
$\tilde{\tau}_{0}(0,t)$ is obviously separable but not so obviously the last separable state on
 $[D_{0},\rho(0,t)]$.  We
confirm that property later. To see if $\tau _{0}\left( 0,t\right) $ lies in the same
face as $\tilde{\tau}_{0}\left( 0,t\right) $, we take normalized
combinations of the four $P_{jkl}^{\pm }$ in the equations above and
minimize the distance to $\rho \left( 0,t\right), $ finding the separable density with four vector 
\[
\left\langle -\frac{t}{2},\frac{t}{2},\frac{1}{8}-\frac{t}{2},\frac{1}{8}-%
\frac{t}{2}\right\rangle.
\]
Using this for $\tau _{0}\left( 0,t\right) $ we find $c_{0}\left(
0,t\right) =\frac{t}{4}$ and $W_{0}\left( 0,t\right) =\frac{t}{4}W_{0}$
where 
\[
W_{0}=\left( 
\begin{array}{cccccccc}
-1 & 0 & 0 & 0 & 0 & 0 & 0 & -2 \\ 
0 & -1 & 0 & 0 & 0 & 0 & -2 & 0 \\ 
0 & 0 & -1 & 0 & 0 & -2 & 0 & 0 \\ 
0 & 0 & 0 & -1 & 2 & 0 & 0 & 0 \\ 
0 & 0 & 0 & 2 & -1 & 0 & 0 & 0 \\ 
0 & 0 & -2 & 0 & 0 & -1 & 0 & 0 \\ 
0 & -2 & 0 & 0 & 0 & 0 & -1 & 0 \\ 
-2 & 0 & 0 & 0 & 0 & 0 & 0 & -1
\end{array}
\right) 
\]

These heuristics work splendidly, and we also find that $W_{0}$ detects the
entanglement of all $\rho \left( c,d\right) $ with $d<c.$ This is 
illustrated in fig.1 where the separating plane is shown. Of course, 
the geometry is more complicated because the hyperplane 
is not two dimensional. As $t$ decreases to $0$, $\rho(0,t)$ moves to the 
center of the line segment $[P^{+}_{111},P^{-}_{221}]$. As $t$ becomes 
negative ($d>c$), $\tau_{0}$ and $\tilde{\tau}_{0}$ move onto a new 
plane where $P_{212}^{+}$ and $P_{122}^{-}$ replace $P_{212}^{-}$ and 
$P_{122}^{+}$. Recall that $D_{0}$ lies at the center of 
$[P^{+}_{ijk},P^{-}_{ijk}]$. The case $m \neq 0$ is easily visualized.

\begin{proposition}
Let $d<c$. Then $\tau _{0}\left( m,t\right) $ with four vector $\left\langle m-\frac{t}{2},
m+\frac{t}{2},\frac{1}{8}-\frac{t}{2},\frac{1}{8}-\frac{t}{2}\right\rangle $ 
is the closest separable density
to $\rho \left( m,t\right) $.
$W_{0}$ is an entanglement witness for every density in $\left\{
\rho \left( c,d\right) :-1/8\leq d<c\leq 1/8\right\}$.
\end{proposition}

\textit{Proof}: Set $m=0$ and let 
$\pi=\pi_{1}\otimes\pi_{2}\otimes\pi_{3}$ denote any separable projection.
Since $Tr\left( W_{0}\rho \left( 0,t\right) \right) <0$ by construction, it
suffices to confirm that $Tr\left( W_{0}\mu \right) \geq 0.$ Defining $\pi_{k}=|\psi_{k}\rangle 
\langle \psi_{k}|$ where 
\begin{displaymath}
|\psi_{k}\rangle =\left[ \begin{array}{c}
\cos\theta_{k} \\
e^{i\phi_{k}}\sin\theta_{k}
\end{array}
\right],
\end{displaymath}
 we
obtain 
\begin{equation}
2Tr\left( W_{0}\mu \right) =2-\sin \left( 2\theta _{1}\right) \sin \left(
2\theta _{2}\right) \sin \left( 2\theta _{3}\right) C\left( \varphi
_{_{1}},\varphi _{_{2}},\varphi _{3}\right)   \label{W0comp}
\end{equation}
where 
\begin{eqnarray*}
C\left( \varphi _{_{1}},\varphi _{_{2}},\varphi _{3}\right)  &=&\cos \left(
\varphi _{_{1}}+\varphi _{_{2}}+\varphi _{3}\right) +\cos \left( \varphi
_{_{1}}+\varphi _{_{2}}-\varphi _{3}\right)  \\
&&+\cos \left( \varphi _{_{1}}-\varphi _{_{2}}+\varphi _{3}\right) -\cos
\left( \varphi _{_{1}}-\varphi _{_{2}}-\varphi _{3}\right) .
\end{eqnarray*}
The phase angles $\varphi _{_{k}}$ can take any value while $0\leq \theta
_{k}\leq \pi /2$. Confirming that the right side of eq.(\ref{W0comp}) is
non-negative is a familiar Bell-inequality computation and proves the
assertion when $m=0$. It follows from comments after 
eq.(\ref{innerproduct}) that 
$\tilde{\tau}_{0}\left( 0,t\right) $ has to lie in the separating plane and
is thus the closest separable state to $\rho \left( 0,t\right) $ along  
$[D_{0},\rho \left( 0,t\right)] $, justifying the notation and the
assumption made earlier.

The generalization to non-zero $m$ is straight-forward. Using the asserted
form for $\tau _{0}\left( m,t\right) $, it's easy to check that $\tau
_{0}\left( m,t\right) $ is separable, that $\rho \left( m,t\right) -\tau
_{0}\left( m,t\right) =\rho \left( 0,t\right) -\tau _{0}\left( 0,t\right) $
and also that $c_{0}\left( m,t\right) =\frac{t}{4}$. It follows that $\tau
_{0}\left( m,t\right) $ is the closest separable state to $\rho \left(
m,t\right) $ and $W_{0}\left( m,t\right) =W_{0}\left( 0,t\right) $,
completing the proof. \ $\Box $

From $\rho \left( m,t\right) =(1+8t)\tilde{\tau}_{0}\left( m,t\right)
-8tD_{0}$ and the form of $\tau _{0}\left( m,t\right) $ we can express the
entanglement witness in terms of the identity and explicit separable states: 
\[
W_{0}\left( 0,t\right) =t\left[ \frac{5}{4}I-2\left(
P_{111}^{+}+P_{221}^{-}+P_{212}^{-}+P_{122}^{+}\right) \right] .
\]
Again we have shown that local detection of entanglement can be defined
using the explicit representations of $\tau _{0}$ and $\tilde{\tau}_{0}$ as
convex combinations of separable projections.

\section{Generalization to $n$ qubits}

In ref. \cite{pitrub2} $\tilde{\tau}_{0}$ was computed for the 
$\rho_{0}$ generated from the $n$-qubit GHZ state defined by
\begin{displaymath}
\rho_{0}=|\psi_{0}\rangle \langle \psi_{0}| \quad {\textstyle where} 
\quad 
|\psi_{0}\rangle=\frac{1}{\sqrt{2}}\left(|\tilde{0}\rangle+|\tilde{1}\rangle \right),
\end{displaymath}
and $\tilde{j}=(j,\cdots,j)$. It was shown that 
\begin{eqnarray}
	\tilde{\tau}_{0} & = & \left(1-s_{0}\right)D_{0}+s_{0}\rho_{0}
	\label{taun}  \\
	 & = & s_{0}\Delta+(1-s_{0})Q
	\label{Q}
\end{eqnarray}
 where 
$s_{0}=1/ \left(2^{n-1}+1\right)$, $\Delta=\frac{1}{2} \left( |\tilde{0}\rangle \langle 
\tilde{0}|+|\tilde{1}\rangle \langle \tilde{1}|\right)$, and $Q$ is a matrix $2^{n}\times2^{n}$ 
matrix  with entries 
$1/2^{n}$ on the diagonal and in the upper and lower corners. It is 
clear that $\Delta$ is a convex combination of two separable states. In 
\cite{pitrub2} $Q$ was expressed in terms of $2^{n-1}$ separable states.
 In \cite{pitrubmat} we also 
computed $\tau_{0}$ and it can be shown that $\tau_{0}$ can be 
expressed as a convex combination of $\Delta$ and $Q$. This is another 
example of a case when $\tau_{0} \neq \tilde{\tau}_{0}$ but both 
densities lie on the near face. Applying Theorem 1 the optimal 
entanglement witness can be written as 
\begin{equation}
W_{0}=aI-b\Delta-cQ
\label{th1W}
\end{equation}
with $a,b,$ and $c$ positive. In the two qubit case this result 
reduces to eq.(\ref{W0}).

\section{Far Face Constructions}

There are cases when an entanglement witness can be defined in terms of the
identity and a separable state without computing the nearest separable
density explicitly. In \cite{bennett} a technique is described for the
construction of inseparable densities with positive partial transposes,
using orthogonal unextendible product bases (\textit{UPB}). This clever
approach assumes a set of $m$ separable orthonormal states $B=\left\{ \left|
\varphi _{k}\right\rangle ,\text{ }1\leq k\leq m\right\} $ where each $
\left| \varphi _{k}\right\rangle $ is a tensor product of states in their
respective Hilbert spaces and where the orthogonal space $B^{\bot }$
contains no separable projections. If $\mu _{k}=\left| \varphi
_{k}\right\rangle \left\langle \varphi _{k}\right| $ and one defines 
\begin{equation}
\mu _{0}=\frac{1}{m}\sum_{k=1}^{m}\mu _{k},  \label{mu0def}
\end{equation}
then 
\[
\rho _{0}=\frac{N}{N-m}D_{0}-\frac{m}{N-m}\mu _{0}
\]
can be shown to be an inseparable density with positive partial transform. A
number of examples of orthogonal \textit{UPB}s are given in \cite{bennett}
and in subsequent papers such as \cite{divenc1} and \cite{divenc2}. The
ideas in \cite{bruss1} also apply in this context and are illustrated there
using the two qutrit example ``TILES'' of \cite{bennett}.

In \cite{pitt} some consequences of the geometric structure implicit in this
approach are developed. For example, it is clear from the equation 
above that $D_{0}$ lies on the line segment 
$[\mu_{0},\rho_{0}]$. If one denotes by $F_{0}$ the face of the
separable densities $S$ containing $\mu _{0}$, then, in the context of the
real Hilbert space $M$, $F_{0}$ is orthogonal to that line. It is 
shown in \cite{terhal} by a compactness argument that there is a positive $\epsilon 
$ such that 
\[
0<\frac{\epsilon }{m}=\inf \left\{ Tr\left[ \mu _{0}\sigma \right] ,\text{ }%
\sigma \in S\right\} 
\]
and thus that the face $G_{0}\equiv \left\{ \sigma \in S:Tr\left[ \mu
_{0}\sigma \right] =\frac{\epsilon }{m}\right\} $ is non-empty. In 
this context it is shown in \cite{pitt} that  
\[
0<s_{0}\equiv 1-\frac{\epsilon N}{m}<1.
\]
A consequence of this approach is that a separating witness $W_{0}$ for 
$\rho _{0}$ can be defined using (\ref{entwitdef}) with 
\[
\tau _{0}=\left( 1-s_{0}\right) D_{0}+s_{0}\rho _{0}.
\]
In this construction $\tau _{0}$ is not necessarily separable but is defined by the intersection
of $[\mu_{0},\rho_{0}]$ and a hyperplane
containing the ``near face'' $G_{0}$. Since both $\rho _{0}$ and $\tau _{0}$
can be written explicitly in terms of $\mu _{0}$, which is separable, then
once $\epsilon $ is known we can again express the entanglement witness in
terms of the identity $I$ and a separable density whose convex
representation is known: 
\[
W_{0}=\frac{\epsilon N}{N-m}\left( \mu _{0}-\frac{\epsilon }{m}I\right) 
\]
Thus, the required coordinated local measurements are defined explicitly by
the original set $B$ and there will be no more than $m$ different settings.
Geometrically $W_{0}$ is expressed in terms of the identity and $\mu _{0}$,
which lives in the far face $F_{0}$, on the ``other side'' of $D_{0}$ from $%
\rho _{0}.$ In the special case discussed in \cite{bruss1}, this is the same
witness as derived there, up to a multiplicative constant.

Consider separable densities $\mu _{b}=\sum_{k}p_{k}\mu _{k}$ in the
face $F_{0}$ that are also close to $\mu _{0}$. Let $b$ denote the
reciprocal of the largest of the coefficients $p_{k}$. Then it is easy to
define inseparable densities 
\[
\rho _{b}=\frac{ND_{0}-b\mu _{b}}{N-b} 
\]
with positive partial transposes that are on the boundary of the set of
densities $D$ and are close to $\rho _{0}$. Moreover $W_{0}$ can
also serve as an entanglement witness for these densities. In fact, using
the same notation as above, one can get a 
``frustram'' of states of the form 
\[
\rho =\left( 1-p\right) \sigma +p\rho _{b} 
\]
which lie in $D$ on the $\rho _{0}$ side of the hyperplane defined by $W_{0}$,
provided 
\begin{equation}
\frac{p\left( m-b\right) }{N-b}+\frac{1-p}{N}+\frac{\left( 1-p\right) \delta 
}{\sqrt{m}}<\frac{\epsilon }{m}  \label{sepeqn2}
\end{equation}
where $\left\| \sigma -D_{0}\right\| <\delta $. We omit the details,
repeating instead that the Euclidean geometry of $M$ provides an extremely
useful context for examining questions of this sort and that the use of 
eq.(\ref
{entwitdef}) gives a unifying geometric approach for constructing
entanglement witnesses.

We should note that the effects of dropping the hypothesis that the states
in $B$ are orthogonal is also discussed in \cite{pitt}, and weaker
conditions on the states in $B$ are given which allow the construction above
of inseparable states to be generalized. In particular, one can perturb the
orthogonal \textit{UPB} case, losing orthogonality but preserving enough of
the structure to allow the analysis to go through. The cost of this
generalization, however, is that the resulting states do not automatically
have positive partial transposes.

\section{Summary}

In this paper we have used a geometric definition of an entanglement witness 
$W_{0}$ detecting an inseparable state $\rho _{0}$ to show that $W_{0}$
always has a representation leading to entanglement detection using
coordinated local measurements. This approach gives essentially the same
witnesses and the same coordinated local measurements as derived in \cite
{bruss1} for their particular two qubit case. When coupled with the
generalized ``spin'' matrices defined in \cite{pitrub1}, it also achieves
the lower bound asserted in \cite{bruss1} for the number of coordinated
local measurements for the analogous $d\times d$ case, at least when $d$ is prime. We
also illustrated the use of the geometry by applying the methodology to a two
parameter family of three qubit bound entangled states for which $\tau _{0}$
and $\tilde{\tau}_{0}$ differ. The strength of the geometrical 
approach is further illustrated by applying it to the $n$ qubit case. In the case of inseparable densities
constructed using orthogonal \textit{UPBs, }the geometric approach also
applies, but produces a representation using a ``far face'' separable density. 

\section{Appendix}

By definition the $S_{\left( j,k\right) }$ ``spin'' matrix is defined as
\[
S_{\left( j,k\right) }=\sum_{r=0}^{d-1}\eta ^{jr}\left| r\right\rangle
\left\langle r+k\right| 
\]
where addition is modulo $d$ and $\eta =\exp \left( 2\pi i/d\right) $. If $u$
denotes $\left( j,k\right) $, then $S_{u}$ has trace $0$ unless $u$ equals $%
e\equiv \left( 0,0\right) $. Orthogonality follows from 
\begin{eqnarray*}
Tr\left[ S_{\left( j_{1},k_{1}\right) }^{\dagger }S_{\left(
j_{2},k_{2}\right) }\right]  &=&Tr\left[ \sum_{r}\sum_{s}\eta ^{-j_{1}r}\eta
^{j_{2}s}\left( \left| r+k_{1}\right\rangle \left\langle r\right| \right)
\left( \left| s\right\rangle \left\langle s+k_{2}\right| \right) \right]  \\
&=&Tr\left[ \sum_{r}\eta ^{\left( j_{2}-j_{1}\right) r}\left|
r+k_{1}\right\rangle \left\langle r+k_{2}\right| \right]  \\
&=&d\delta_{j_{1},j_{2}}\delta_{k_{1},k_{2}}.
\end{eqnarray*}
Similarly, one can calculate some useful relations such as $%
S_{0,1}S_{1,0}=\eta S_{1,0}S_{0,1}$, $S_{j,k}=\left( S_{1,0}\right)
^{j}\left( S_{0,1}\right) ^{k}$, $\left( S_{j,k}\right) ^{m}=\eta
^{jkm\left( m-1\right) /2}S_{mj,mk}$, and $S_{j,k}^{\dagger }=\eta
^{jk}S_{d-j,d-k}.$ Unlike the Pauli matrices, the $S_{u}$ are not
necessarily Hermitian, but they are unitary and can play a role analogous to
that played by the Pauli matrices.

Any $d\times d$ density $\alpha $ can thus be written as a linear
combination of these spin matrices, and we have 
\[
\alpha =\frac{1}{d}\left[ S_{e}+\sum_{u\neq e}s_{u}S_{u}\right] 
\]
where we use $u=\left( j,k\right) $ in 
\[
s_{u}=Tr\left[ S_{u}^{\dagger }\alpha \right] =\sum_{r}\eta ^{-jr}\alpha
_{r,r+k}. 
\]
To represent a density such as $\tau _{0}$ defined in eq.(\ref{dtau}) on the tensor product
space $H^{\left[ d\right] }\otimes H^{\left[ d\right] }$, we use the set of
tensor products of the spin matrices as an orthogonal basis. A direct
calculation or an invocation of eq.(16) of ref. \cite{pitrub1} gives 
\[
\tau _{0}=\frac{1}{d^{2}}\left[ \frac{d}{d+1}S_{00,00}+\frac{1}{d+1}%
\sum_{k=0}^{d-1}\sum_{i=0}^{d-1}S_{i(d-i),kk}\right] 
\]
where $S_{ij,kl}=S_{i,k}\otimes S_{j,l}$. It is at this point that we 
require $d$ be prime. Then for given $i$
and $k\neq 0$ there is a unique $j$ such that $i=jk$ (mod $d$), and we can rewrite $\tau _{0}$ in the form 
\begin{eqnarray*}
\tau _{0} &=&\frac{1}{d+1}\left[ \frac{d}{d^{2}}S_{00,00}+\frac{1}{d^{2}}
\sum_{i=0}^{d-1}S_{i(d-i),00}+\frac{1}{d^{2}}\sum_{k=1}^{d-1}%
\sum_{j=0}^{d-1}S_{(kj)(kd-kj),kk}\right] \\
&=&\frac{1}{d+1}\left[ \frac{1}{d^{2}}\sum_{k=0}^{d-1}S_{k(d-k),00}+%
\sum_{j=0}^{d-1}\left( \frac{1}{d^{2}}\sum_{k=0}^{d-1}S_{(kj)(kd-kj),kk}%
\right) \right].
\end{eqnarray*}

It remains to show that each of the expressions involving a $k$-summation is
a summation of tensor products of projections from a complete set of
orthogonal projections. That is, each summation corresponds to correlated
local measurements, and $\tau _{0}$ is realized by $d+1$ such summations.

We begin by defining a complete set of projections in terms of the spin
matrices, a construction which corresponds to that in the spin $1/2$ context.

\begin{lemma}
(Reference \cite{pitrub1}) Let $d>2$ be prime and let $e\neq u=\left( j,k\right) $.
Then if 
\[
P_{u}\left( r\right) \equiv \frac{1}{d}\sum_{m=0}^{d-1}\left( \eta
^{r}S_{u}\right) ^{m}=\frac{1}{d}\sum_{m=0}^{d-1}\eta ^{mr}\eta ^{jkm\left(
m-1\right) /2}S_{mu},
\]
$\left\{ P_{u}\left( r\right) :0\leq r<d\right\} $ is a complete set of
trace one, orthogonal (Hermitian) projections.
\end{lemma}

\textit{Proof}: $P_{u}\left( r\right) $ has trace one since the only term
with non-zero trace is the $m=0$ term. From the definition 
\begin{eqnarray*}
P_{u}\left( r\right) P_{u}\left( s\right) &=&\frac{1}{d^{2}}%
\sum_{m=0}^{d-1}\sum_{n=0}^{d-1}\eta ^{mr+ns}\eta ^{jk\left[ m\left(
m-1\right) +n\left( n-1\right) \right] /2}S_{mu}S_{nu} \\
&=&\frac{1}{d^{2}}\sum_{m=0}^{d-1}\sum_{n=0}^{d-1}\eta ^{\left( m+n\right)
r}\eta ^{jk\left[ m\left( m-1\right) +n\left( n-1\right) \right] /2}\eta
^{jkmn}S_{\left( m+n\right) u}\eta ^{n\left( s-r\right) }.
\end{eqnarray*}
Make the substitution $t=m+n$ in the last expression and collect terms to
obtain 
\[
P_{u}\left( r\right) P_{u}\left( s\right) =\left( \frac{1}{d}%
\sum_{n=0}^{d-1}\eta ^{n\left( s-r\right) }\right) \frac{1}{d}%
\sum_{t=0}^{d-1}\eta ^{tr}\eta ^{jkt\left( t-1\right) /2}S_{tu}, 
\]
thereby obtaining both the orthogonality and $P_{u}\left( r\right)
P_{u}\left( r\right) =P_{u}\left( r\right) .$ Finally 
\begin{eqnarray*}
\left( P_{u}\left( r\right) \right) ^{\dagger } &=&\frac{1}{d}%
\sum_{m=0}^{d-1}\eta ^{-mr}\left( S_{u}^{\dagger }\right) ^{m}=\frac{1}{d}%
\sum_{m=0}^{d-1}\eta ^{-mr}\left( \eta ^{jk}S_{-u}\right) ^{m} \\
&=&\frac{1}{d}\sum_{m=0}^{d-1}\eta ^{-mr+mjk+jkm(m-1)/2}S_{-mu}=\frac{1}{d}%
\sum_{n=0}^{d-1}\eta ^{nr+jkn(n-1)/2}S_{nu} \\
&=&P_{u}\left( r\right) .
\end{eqnarray*}
These steps actually introduce a factor of the form $\eta ^{jkd\left(
d-1\right) /2}$ which equals $1$ for odd integers. However, if $d$ is even
and $j\ $and $k$ are odd, $\eta ^{jkd\left( d-1\right) /2}\neq 1$, and the
proof must be modified. $\Box $

Having defined complete sets of projections, we are ready for the final
technical result.

\begin{proposition}
Let $u_{j}=(j,1)$ and $v_{j}=\left( d-j,1\right) $for $0\leq j<d.$ Then 
\[
\frac{1}{d}\sum_{r=0}^{d-1}P_{u_{j}}\left( r\right) \otimes P_{v_{j}}\left(
d-r\right) =\left( \frac{1}{d^{2}}\sum_{k=0}^{d-1}S_{(kj)(kd-kj),kk}\right) .
\]
If $x=\left( 1,0\right) $ and $y=\left( d-1,0\right) ,$ then 
\[
\frac{1}{d}\sum_{r=0}^{d-1}P_{x}\left( r\right) \otimes P_{y}\left(
d-r\right) =\frac{1}{d^{2}}\sum_{k=0}^{d-1}S_{k(d-k),00}.
\]
\end{proposition}

\textit{Proof}: The proof is just a matter of navigating the notation.
Suppressing the subscript, 
\begin{eqnarray*}
\frac{1}{d}\sum_{r=0}^{d-1}P_{u}\left( r\right) \otimes P_{v}\left(
d-r\right) &=&\frac{1}{d^{3}}\sum_{k,n}\left( S_{u}\right) ^{k}\otimes
\left( S_{v}\right) ^{n}\sum_{r}\eta ^{\left( k-n\right) r} \\
&=&\frac{1}{d^{2}}\sum_{k}S_{ku}\otimes S_{kv}\eta ^{\left[ jk\left(
k-1\right) /2+\left( d-j\right) k\left( k-1\right) /2\right] } \\
&=&\frac{1}{d^{2}}\sum_{m}S_{ku}\otimes S_{kv}=\frac{1}{d^{2}}
\sum_{k=0}^{d-1}S_{(kj)(kd-kj),kk}
\end{eqnarray*}
as required. The proof of the remaining assertion is similar, and we omit
the details. $\Box $

\pagebreak

\begin{figure}[t]
\centerline{
\input epsf
\setlength{\epsfxsize}{4in}
\epsffile{fig1.eps}
}
\end{figure}

\begin{figure}
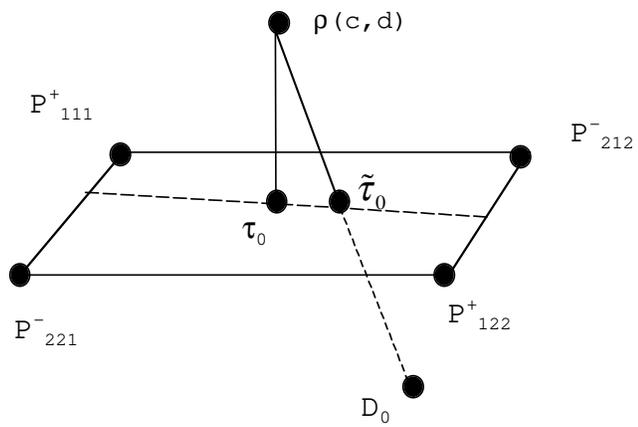

\caption{The plane containing $\tau_{0},\tilde{\tau}_{0}$, and the 
four separable densities of eq. (\ref{tautilde}) is illustrated. The separable states lie 
below the plane and the random state $D_{0}$ is shown.}
\label{fig1.eps}
\end{figure}


\begin{thebibliography}{99}

\bibitem{pitrubmat}  A. O. Pittenger, M. H. Rubin, ``Convexity and the
separability problem of quantum mechanical density matrices,'' Lin. Alg.
Appl. \textbf{346} (75 - 91), (May 2002).


\bibitem{bruss1}  O. G\"{u}hne, P. Hyllus, D. Bruss, A Ekert, M. Lewenstein,
C. Macchiavello, A. Sanpera, ``Detection of entanglement with few local
measurements'', quant-ph/0205089 (May 2002).

\bibitem{pitt}  A. O. Pittenger, ``Unextendible product bases and the
construction of inseparable states,'' Lin. Alg. Appl., to appear.

\bibitem{bell}  J. Bell, \textit{Speakable and Unspeakable in Quantum
Mechanics}, Cambridge Univ. Press, London and New York, (1993).

\bibitem{terhal0}B. M. Terhal, Phys. Lett. A{\bf271}, 319 (2000).

\bibitem{ekhor}  Artur Ekert, Pawel Horodecki, ``Direct detection of quantum
entanglement'', quant-ph/0111064(Nov. 2001).

\bibitem{horodeckis3}  M. Horodecki, P. Horodecki, R. Horodecki,
``Separability of mixed states: necessary and sufficient conditions,'' Phys.
Lettrs. A, \textbf{223} 1-8, (1996), quant-ph/9605038.

\bibitem{terhal}  B. M. Terhal, ``Detecting quantum entanglement,''
quant-ph/0101032 (Jan. 2001).


\bibitem{lewenstein}  M. Lewenstein, B. Kraus, J. I. Cirac, P. Horodecki,
``Optimization of entanglement witnesses,'' quant-ph/0005014 (Jan. 2000).

\bibitem{zuk}  K. Zyczkowski, P. Horodecki, A. Sanpera, M. Lewenstein, ``On
the volume of mixed entangled states,'' Phys. Rev. A \textbf{58}, 883 (1998).

\bibitem{pitrub3}  A. O. Pittenger, M. H. Rubin, ``Note on the separability
of the Werner states in arbitrary dimensions,'' Optics Comm. \textbf{179},
447 - 449 (2000) and quant-ph/0001110 (Jan. 2000).

\bibitem{steiner1}  R. B. Lockhart, M. J. Steiner, ``Preserving entanglement
under decoherence and sandwiching all separable states'',
quant/ph 0009090 (Sep. 2000).

\bibitem{steiner2}  R. B. Lockhart, M. J. Steiner, K. Gerlach, ``Geometry
and product states'',  quant/ph 0010013 (Oct. 2000).

\bibitem{witru}  C. Witte, M. Trucks, ``A new entanglement measure induced
by the Hilbert-Schmidt norm'', Phys.Lett. A 257, 14-20, (1999).


\bibitem{pitrub1}  A. O. Pittenger, M. H. Rubin, ``Separability and Fourier
representations of density matrices,'' Phys. Rev. A, \textbf{62} 032313
(2000).

\bibitem{fivel}  D. I. Fivel, ``Remarkable phase oscillations appearing in
the lattice dynamics of Einstein-Podolsky-Rosen states,'' Phys. Rev. Lett. 
\textbf{74}, 835 (1995).

\bibitem{pitrub2}  A. O. Pittenger, M. H. Rubin, ``Complete separability and
Fourier representations of $n$-qubit states,'' Phys. Rev. A, \textbf{62}
042306 (2000).

\bibitem{bennett}  C. H. Bennett, D. P. DiVincenzo, T. Mor, J. A. Smolin, B.
M. Terhal, ``Unextendible product bases and bound entanglement'', Phys. Rev.
Lett. \textbf{82}, 5385 (1999).

\bibitem{divenc1}  D. P. DiVincenzo, T. Mor, P. W. Shor, J. A. Smolin, B. M.
Terhal, ``Unextendible product bases, uncompletable product bases and bound
entanglement'', quant-ph/9908070 (Nov 2000).

\bibitem{divenc2}  D. P. DiVincenzo, B. M. Terhal, ``Product bases in
quantum information theory'', quant-ph/9008055 (Aug 2000).


\end{thebibliography}
\end{document}